\documentstyle[12pt,psfig]{article}

\newcommand{\ra}{\rightarrow}

\newcommand{\be}{\begin{equation}}
\newcommand{\ee}{\end{equation}}

\newcommand{\bea}{\begin{eqnarray}}
\newcommand{\eea}{\end{eqnarray}}
\newcommand{\bef}{\begin{figure}}
\newcommand{\eef}{\end{figure}}

\begin{document}
\vskip 0.2in
\begin{center}

{\large{\bf Charmonium Productions: Nucleon-Nucleon versus Nucleus-Nucleus
Collisions}}
\end{center}
\vskip 0.1in
\begin{center}
{\normalsize 
Pradip Roy$^a$, Abhee K. Dutt-Mazumder$^a$, 
Jan-e Alam$^b$
}
\vskip .1in
a) Saha Institute of Nuclear Physics, 1/AF Bidhan Nagar, Kolkata, India
\\
b) Variable Energy Cyclotron Centre, 1/AF Bidhan Nagar, Kolkata, India
\\
\vskip .1in
\end{center}

\addtolength{\baselineskip}{0.4\baselineskip} 

\parindent=20pt
\vskip 0.1in
\begin{abstract}
Charmonium productions in $p-p$ and $A-B$ collisions have been estimated
within the ambit of colour evaporation model (CEM). The model parameters 
have been fixed by fitting the theoretical results with CDF data.
The method is then applied to RHIC and LHC energies to obtain the 
transverse momentum distributions of $J/\psi$, $\psi^{\prime}$ and 
$\chi_c$. Suppression due to Debye screening in a quark gluon plasma 
(QGP) is estimated at various centrality cuts. The final $p_T$ distributions 
of various resonances are then predicted convoluting with the survival 
probability.
\end{abstract}

\section*{I. Introduction}
Ever since the possibility of creating quark gluon plasma (QGP)
in relativistic heavy ion collision was envisaged, numerous 
signals were proposed to probe the properties of such an exotic
state of matter.  In this context 
Satz and Matsui~\cite{satz} had suggested that
the production of heavy quark resonances ($J/\psi$) will be suppressed 
as a result of colour Debye screening in a hot
and dense system of quarks, anti-quarks and gluons. 
This suppression could
be detected experimentally through the dileptonic decay mode of these
resonances. ALICE dimuon spectrometer~\cite{muonlab} is dedicated to look 
for this
type of signal. However, it is a daunting task to 
disentangle  the contributions of the heavy quarkonium states
to muon spectrum due to the 
background from several other sources, {\it e.g.}
Drell-Yan, semileptonic
decay of open heavy flavoured mesons ($D{\bar D}, B{\bar B}$) etc. 
Low energy muons
from kaons and pions also constitute a large background. 


In this work we shall estimate the hard charmonium productions 
({\it i.e.} $J/\psi$ 
produced from initial hard process, will
be called hard $J/\psi$ hereafter) both in $p-p$ and $A-A$ collisions.
The production of heavy resonances
 proceeds 
via the following two steps:
(i) the production of heavy quark-antiquark pairs (perturbative), 
(ii) their resonance interactions to form the bound state (non-perturbative).

The initial state in relativistic heavy ion collisions 
consists of either hadronic matter  
or QGP depending on the incident energies
of the colliding nuclei. At LHC energies
the formation of QGP is unavoidable.
Even if the system is formed in QGP phase it
will revert back to hadronic phase due to the cooling
of the expanding QGP system
and hence the interaction of the hard $J/\psi$ 
with the hadronic
matter is inevitable. Therefore, in addition to the suppression due
to Debye screening, one needs
to consider the survival probability of those
$J/\psi$ due to its interactions with the hadronic
medium. In this work we have neglected the suppressions due to 
co-movers as this effect is found to be negligibly small.


The paper is organized as follows.
In section II we shall describe the formalism for charmonium  productions
in the CEM along with the survival probability as a function of centrality. 
In section III results of our calculations will
be presented followed by summary and discussion in section IV.

\section*{II. Charmonium Productions}

\subsection*{a. Nucleon-Nucleon Collisions}

The CEM (also known as local duality 
approach)~\cite{schuler,kramer,mangano,cem,schuler1,gavai}
states that a heavy quark pair
with mass $M_{{Q \bar Q}} < 2 M_{Q {\bar q}}$ transforms ($q$ designates light
quark, while $Q$ stands for $c$ or $b$ quarks ), independent to its
colour and spin, to a $Q{\bar Q}$ bound state. The production of a particular
bound state depends on the dynamical details of hadronization procedure.
The bound state formation probability ($F[nJ^{PC}]$;  the so called 
normalization factor) cannot be calculated from 
first principle. This is treated as a parameter and can be extracted by fitting
the model with the experimental data. $F[nJ^{PC}]$  depends on the 
particular resonance state under consideration,  mass of the heavy flavour, 
the order (LO or NLO) of the $Q\overline{Q}$ production and the type
of parton distribution function (PDF) used. Another important point
is that the produced $Q{\bar Q}$ is not constrained to have the proper 
spin-parity and colour neutrality. The pair sheds its colour non-perturbatively
to evolve into asymptotic state without affecting the cross-section.
This model cannot predict the absolute cross-section of heavy resonance
production, it however, predicts, their $p_T$ and $\sqrt{s}$ dependence.
As mentioned, CEM does not bother about the spin-parity and colour
neutrality of the bound state. This constraint can be removed by considering
an improved version of the duality approach, where one assumes that only
the colour singlet part of the cross-section contributes to the bound 
state production. It might be recalled here that the CEM contains
some features of non-relativistic QCD, due to the inclusion of colour
octet processes. 

In spite of these limitations of CEM,
it is capable of explaining $p_T$ distribution of charmonium in hadron-hadron
collisions reasonably well.  As mentioned before, the production of 
charmonium consists of two stages; production of a $c{\bar c}$ pair 
(perturbative process) and subsequent non-perturbative evolution into 
asymptotic states.  We have considered those
hard processes which can contribute to $c\bar{c}$ productions irrespective
of their colour and spin-parity. The colour 
neutralization occurs by the interactions (one or more soft gluon emission)
with the surrounding colour fields and this step is considered to be 
non-perturbative. 
In CEM quarkonium production is treated identically to open heavy flavour
production with the exception that in the case of quarkonium, the invariant
mass of the heavy quark pair is restricted below the open charm/bottom 
mesons  threshold
(see eq.~(\ref{eq2}) below),
which is twice the mass of the lowest meson mass that can be formed with 
the heavy quark.  Depending on the quantum numbers of the $Q{\bar Q}$
pair different matrix elements are needed for various resonances. 
The effects of these non-perturbative matrix
elements are combined into the factor $F[nJ^{PC}]$ which is 
universal~\cite{FnJ} {\it i. e.} process and kinematics independent. 
It describes the probability that the
$Q{\bar Q}$ pair forms a quarkonium of given spin ($J$), parity ($P$)
and charge conjugation ($C$). The production cross section for a $J/\psi$ or
any charmonium state is therefore given by~\cite{FnJ}
\be
\sigma[R(nJ^{PC})] = F[nJ^{PC}]\,{\tilde {\sigma}}[Q{\bar Q}]
,
\label{eq1}
\ee  
where the non-perturbative (long distance) factor can be written 
in terms of the probability
to have colour singlet state (1/9) and the fraction $\rho_R$ of each specific
charmonium state. The perturbative contribution (short distance) 
is given by
\be
{\tilde{\sigma}}[Q{\bar Q}] = \int_{2m_Q}^{2m_{D/B}}\,dM_{Q{\bar Q}}^2
\,\frac{d\sigma[Q{\bar Q}]}{dM_{Q{\bar Q}}^2}.
\label{eq2}
\ee

The contributions to heavy quark production in leading order come from
$q\,{\bar q}\,\ra\,Q\,{\bar Q}$ and $g\,g\,\ra\,Q\,{\bar Q}$. The differential
cross-section for heavy quark pair production in hadron-hadron 
collision is given by~\cite{vogt}
\be
\frac{d\sigma}
{dM_{Q{\bar Q}}^2 dy}
[h_A h_B\ra Q {\bar Q}X] = 
\frac{H(x_a,x_b,Q^2)}{s},
\label{eq3}
\ee
where 
\bea
H(x_a,x_b,Q^2)& =& \sum_f\,\left[\frac{}{}q_f^{h_A}(x_a,Q^2)
{\bar q}_f^{h_B}(x_b.Q^2)+\frac{}{}{\bar q}_f^{h_A}(x_a,Q^2)
q_f^{h_B}(x_b,Q^2)\right]{\hat {\sigma}}_{q{\bar q}\ra Q{\bar Q}}\nonumber\\
&&+g^{h_A}(x_a,Q^2)g^{h_B}(x_b,Q^2){\hat {\sigma}}_{gg\ra Q{\bar Q}}.
\label{eq4}
\eea
$x_{a,b} = M_{Q\bar{Q}}\,e^{\pm y}/\sqrt{s}$, $\sqrt{s}$ being the centre of mass
energy of the hadronic system, $y$ stands for rapidity
and $M_{Q\bar{Q}}$ is the invariant mass of the pair. 
$q_f$'s and $g$'s are the PDFs
for quarks and gluons respectively,
these are to be taken from either CTEQ or MRST or GRV~\cite{pdf}. 
Results presented in this work have been obtained with
CTEQ(LO) distribution function.
Combining eqs.(\ref{eq1}),(\ref{eq2}) and (\ref{eq3}) we obtain the 
cross-section for resonance production  per unit rapidity as,
\be
\frac{d\sigma}
{dy}
[h_A h_B\ra R X] = 
F[nJ^{PC}]\,\int_{2m_Q}^{2m_{D/B}}\,dM_{Q{\bar Q}}^2
\frac{H(x_a,x_b,M_{Q{\bar Q}}^2)}{s}.
\label{eq5}
\ee

The above equation can be used to calculate the longitudinal momentum
dependence of the quarkonium production cross section as shown in 
Ref.~\cite{vogt}. 

 The leading order (LO) calculation does not have any $p_T$ dependence
as the heavy quark pairs are produced with $p_T = 0$ (assuming
there is no $p_T$ broadening of the initial state partons). In order to
obtain the $p_T$ distribution one has to go beyond LO. The dominant
production mechanism of heavy quark pairs with large $p_T$ and
invariant mass near the threshold
is the large $p_T$ gluon splitting with the probability given 
by~\cite{mangano}

\begin{equation}
\frac{d{\mathrm{Prob}}}{dM_{Q {\bar Q}}^2} = \frac{\alpha_s}{6\pi}\,
\frac{1}{M_{Q {\bar Q}}^2}
\label{eq6}
\end{equation}
and
\begin{equation}
\frac{d\sigma_{R}^{NN}} {d^2p_Tdy}
 = F[nJ^{PC}]\,\int_{2m_Q}^{2m_{D/B}}\,dM_{Q {\bar Q}}^2\,
\frac{d\sigma^g}{d^2p_Tdy}\,\frac{\alpha_s}{6\pi}\,
\frac{1}{M_{Q {\bar Q}}^2},
\label{eq7}
\end{equation}
where $N$ stands for nucleon,
$d\sigma^g/d^2p_Tdy$ is the inclusive gluon $p_T$ and $y$ distribution 
in $NN$
collisions calculated in LO. It is given by ($a\,b\,\ra\,g\,c$)
\begin{equation}
\frac{d\sigma_g}{d^2p_Tdy}=\frac{1}{16\pi^2s^2}\sum_{a\,b}\,\int\,
G_{a/N}(x_a,Q^2)\,G_{b/N}(x_b,Q^2)\,dy_4\,\frac{{\langle {\cal M}\rangle}^2}
{x_a x_b},
\label{eq8}
\end{equation}
where $x_a = p_T(e^{y}+e^{y4})/\sqrt{s}$,
$x_b = p_T(e^{-y}+e^{-y4})/\sqrt{s}$, ${\langle {\cal M} \rangle}^2$ is the
matrix element for the process and $Gs$ are the 
parton distribution function. $p_T$ and $y$ are the transverse momentum and
rapidity of the gluon which splits into ${Q{\bar Q}}$ pair and $y_4$ is the
rapidity of the particle $c$.
The processes that contribute to the heavy quark pair productions are
$q\,{\bar q}\,\ra\,g\,g$, $g\,g\,\ra\,g\,g$ and $g\,q ({\bar q})\,\ra\,
g\,q ({\bar q})$. Here the final state gluon splits into a heavy pair.
Our analyses show that  $F[J/\psi]$ = 0.045 reproduces the CDF~\cite{cdf} data 
quite well (see later).

\subsection*{b. Nucleus-Nucleus collisions}

The charmonium production, is observed to be suppressed 
both in $p$-$A$ and $A$-$B$ collisions (compared to the scaled $p$-$p$  
scattering).  This could be either due to nuclear absorption
or due to the reduction of  $Q$-$\bar{Q}$ interaction range in a QCD plasma
caused by Debye screening. While the former is known as the normal
nuclear suppression, the latter driven by the plasma effect is dubbed as  
anomalous suppression. 

To understand the mechanism of anomalous 
suppression one introduces the concept of quarkonium formation time 
($\tau_{0f}$)
and the dissociation temperature $T_d$ determined
from the condition at which the $Q$-$\bar{Q}$ interaction range becomes 
equal to the size of the quarkonium. The corresponding time
when plasma attains a temperature $T= T_d$, is denoted as $\tau_d$.
$\tau_{0f}$ on the other hand is the time required for 
the $Q$-$\bar{Q}$ pair to evolve into a physical charmonium state.
This in the plasma rest frame would be Lorentz dilated and reads as 
$\tau_{\mathrm{form}}\equiv \gamma \tau_{0f}=\tau_{0f}\sqrt{1+p_T^2/m_R^2}$.
High $p_T$ quarkonium states can evade suppression
under two circumstances: (i) if $Q-{\bar Q}$ pair materializes into a bound 
state when the plasma has cooled down below $T_d$ 
or (ii) when the bound state is formed outside 
the plasma. The first condition, {\it i. e.} $\tau_{\mathrm{form}} > \tau_d$ 
implies no
suppression for $p_T > p_{1T}^{\mathrm{crit}}$   
while $p_T > p_{2T}^{\mathrm{crit}}$ for no suppression stems
from the condition, $|\vec{r}+\tau_{0f}\vec{p_T}/m_R|>R_s$. 
Here $R_s$ is the radius of the screening zone 
(see later)
and $\vec{r}$ is the position where the heavy quark pair is produced.
Therefore anomalous suppresion will be realized for 
quarkonium momenta less than 
min$\lbrace p_{1T}^{\mathrm{crit}},\,p_{2T}^{\mathrm{crit}}\rbrace$.

Next we consider $J/\psi (\psi^{\prime}, \chi_c)$ production in 
$p-A$ and $A-B$ collisions.
To this end, we first briefly mention the necessary formulae in Glauber 
model~\cite{glau,wong}.
The total inelastic cross-section
in $A-B$ collisions at an impact parameter $b$ is given by
\be
\frac{d\sigma_{\mathrm {in}}^{AB}}{d\bf{b}} = 1-\left[\frac{}{}
1-T_{AB}(\bf{b})\,\sigma_{\mathrm{in}}^{NN}\right]^{AB}
\equiv\,1 - P_0({\bf{b}}),
\label{eq9}
\ee
where $T_{AB}(\bf{b})$ is the nuclear overlap function given by
\be
T_{AB}({\bf{b}}) = \int\,d{\bf{s}}\,T_A({\bf{b}})\,T_B({\bf{b}-\bf{s}})
\label{eq10}
\ee
The nuclear thickness functions are normalized to unity, {\it{i. e.}}
$\int\,d{\bf{b}}\,T_A({\bf{b}}) =\int\,d{\bf{b}}\,dz\,\rho_A({\bf{b}},z) = 1$, 
here $\rho_A$
is the nuclear density distribution.

Generally we are interested in the cross-sections for a set of
events in a given centrality range defined by the trigger settings.
Centrality selection corresponds to a cut on the impact parameter,
$b$ of the collisions. The sample of events in a given centrality
range $0 \leq b \leq b_m$, contains a fraction of the 
total inelastic cross-section.
This fraction is defined by \cite{wong1}
\bea
f(b_m)&=&\frac{
\int_0^{b_m}d{\bf{b}}\,
\frac{d\sigma_{\mathrm {in}}^{AB}}{d\bf{b}}}
{\int_0^\infty\,d\bf{b}\,
\frac{d\sigma_{\mathrm{in}}^{AB}}{d\bf{b}}}
\label{eq11}
\eea

Now we discuss the $J/\psi$ survival probability when, after production, it propagates 
through the target/projectile nucleus.
After creation, the $J/\psi$ meson can interact with other nucleons in the
target and the projectile and may get destroyed mainly due to $J/\psi - N$
interactions. The cross-section for $J/\psi$ production in
$p-A$ collisions can be written as~\cite{khar}
\be
\sigma_{J/\psi}^{pA}(b_m) = A\int_0^{b_m}\,d{\bf{b}}\,dz\,\rho_A({\bf{b}},z)\,
\exp{\left[-(A-1)\int_z^\infty\,\sigma_{\mathrm{abs}}\,\rho_A({\bf{b}},{z^{\prime}})\,
d{z^{\prime}}\right]}\,{{\sigma}}_{J/\psi}^{NN}, 
\label{eq12}
\ee
where ${\sigma}_{J/\psi}^{NN}$ is obtained from 
eq.(\ref{eq5}). The interpretation of 
the above equation is as follows.
The resonance is formed at ${\vec r} = ({\bf{b}},z)$ where the density 
of the target nucleus is $\rho_A({\vec r})$. It can travel in forward
direction ($z$) at constant impact parameter and its intensity is
attenuated due to $J/\psi-N$ inelastic collisions. The exponential
factor accounts for this attenuation. 

The generalization of eq.(\ref{eq12}) in nucleus-nucleus collisions is 
straightforward.
The $J/\psi$ production cross-section in $A-B$ collisions at an
impact parameter $b$ can be written as 
\bea
\frac{d\sigma_{J/\psi}^{AB}}{d^2bdp_T}(b)& = &
\frac{d{{\sigma}}_{J/\psi}^{NN}}{dp_T}\,
AB\,\int\,d{\bf{s}}\,dz_1\,dz_2\,\rho_A(s,z_1)\,
\rho_B({\bf{ b}}-{\bf{s}},z_2)\nonumber\\
&&\times\,\exp{\left(-(A-1)\int_{z_1}^{\infty}\,\sigma_{\mathrm{abs}}\,
\rho_A({\bf{s}},{z^{\prime}})\,
d{z^{\prime}}\right)}\,\nonumber\\
&&\times\,\exp{\left(-(B-1)\int_{z_2}^{\infty}\,\sigma_{\mathrm{abs}}\,
\rho_B({\bf{b}}-{\bf{s}},{z^{\prime}})\,
d{z^{\prime}}\right)}
\label{eq13}
\eea

The observed charmonium suppression data at SPS energies~\cite{spsdata}
could be explained either by the Debye screening in a QGP or 
by co-mover scattering
due to hadronic matter in the initial state~\cite{khar,spsth}. But the co-mover 
interpretation
of heavy quarkonium suppression at LHC energies does not seem to be plausible,
as the initial temperatures achieved at these energies could be considerably
larger than the transition temperature, $T_c$. Thus, the heavy quarkonium
will encounter the hadronic comover at a much later time.
It should be noted here that the formation of QGP does not
necessarily guarantee the suppression of the heavy resonances.
The temperature of the system must be greater than the dissociation
temperature  of the particular resonance in order to get suppressed.
The dissociation temperature
 has recently been calculated using lattice QCD~\cite{lat1,lat2}.
The most recent values of $T_d$ for different heavy flavour resonances
are shown in Table I. Thus as long as the plasma temperature remains greater
than $T_d$ bound state cannot be formed.  
The time $\tau_d$ is obtained by Bjorken scaling law~\cite{bj}:  
\begin{equation}
\tau_d(b) = \tau_i\,\left[\frac{T_i(b)}{T_d}\right]^3,
\label{eq14}
\end{equation}
where $\tau_i$ and $T_i(b)$ (defined later) are the plasma formation 
time and initial temperature of the
plasma respectively. In order to obtain the centrality dependence of 
the suppression we
estimate the initial temperature by assuming the isentropic expansion of the
system, namely,

\vskip .1in
\begin{center}
\begin{tabular}{|lr|cccc|}\hline
& Resonance & $T_d/T_c$~\cite{lat1}& $T_d/T_c$~\cite{lat1}& $\langle T_d\rangle/T_c$&
\\
\hline
&$J/\psi$ & $1.1$&$2.0$ & 1.5 &
\\
\hline
&$\chi_c$  & $0.74$& $1.1$ & 0.9 &
\\
\hline
&$\psi^{\prime}$  & $0.1 - 0.2$& $1.1$ & 0.625 &
\\
\hline
&$\Upsilon$  & $2.31$& $4.5$ & 3.4 &
\\
\hline
&$\chi_b$  & $1.13$& $2.0$ & 1.55 &
\\
\hline
&$\Upsilon^{\prime}$  & $1.1$& $2.0$ & 1.55 &
\\
\hline
&$\chi_b^{\prime}$  & $0.83$& $-$ & $-$ &
\\
\hline
&$\Upsilon^{\prime \prime}$  & $0.74$& $-$ & $-$ &
\\
\hline
\end{tabular}
\end{center}
{Table I: Dissociation temperatures of charmonium and 
bottomonium system in pure gluonic plasma.}
\vskip .1in
 
\begin{equation}
T_i^3(b)= \frac{2\pi^4}{45 \zeta(3)}\,\frac{1}{4a_i \pi R_t^2(b) \tau_i}\,
\frac{dN_{AB}}{dy}(b).
\label{eq15}
\end{equation}
$a_i =\pi^2/90(21N_F/2+16)$, is the degeneracy in the initial state,
$dN_{AB}/dy$ is the total multiplicity of produced hadrons measured 
experimentally and $R_t$ is the transverse dimension of the plasma.. 
One would expect that the total multiplicity
in nucleus-nucleus collisions at RHIC and LHC energies will come from both
soft as well as hard 
collisions. It has been shown recently ~\cite{nardi} 
that about 10 \% of the total multiplicity at RHIC energies
 comes from hard collisions.
Therefore, the total hadron multiplicity in nucleus-nucleus collisions at
a given impact parameter can be written as (the so called two-component model)
\be
\frac{dN_{AB}}{dy}(b) = \left[(1-x)\,\frac{N_{\mathrm part}(b)
}{2}
+ x\,N_{\mathrm coll}(b)\right]\,\frac{dN_{NN}}{dy},
\label{eq16}
\ee
In the above $x$ is the fraction of hard collisions.
For hadron rapidity density ($dN_{NN}/dy$) in nucleon-nucleon collision we use
the following form~\cite{abe}: $dN_{NN}/dy = 
1.5(2.5 - 0.25\,\ln(s) +0.023\,\ln^2(s))$.
$N_{\mathrm{part}}(b)$ and
$N_{\mathrm{coll}}(b)$ are the  numbers of participants
and number of collisions  respectively at an impact parameter $b$. 
In non-central collisions the plasma transverse radius can be 
approximated as $R_t\sim$ 1.2[0.5$N_{\mathrm{part}}(b)]^{1/3}$\cite{marek}. 
The average initial temperatures and the dissociation times 
for various centrality classes have been shown in Table II. 

The critical radius $R_s$, beyond which there will be no suppression, is
given by~\cite{chum}
\be
R_s = R_t\,\left[1-\left(\frac{\gamma \tau_{0f}}{\tau_d}\right)^4\right]^{1/2}
\label{eq17}
\ee

\begin{center}
\begin{tabular}{|lr|c|c|c|c|}\hline
& Centrality (fm) &$R_t$ (fm)& $\langle T_i\rangle$ (GeV)  & $\langle\tau_d\rangle$ (fm) &
\\
\hline
& $[0-3]$ &6.9     &  0.617 &   4.43 & 
\\
\hline
 & $[3-6]$  &6.36    &  0.582 &  3.73  & 
\\
\hline
 & $[6-9]$ &5.38     &  0.522 &  2.70  & 
\\
\hline
 & $[9-12]$ &4.0     &  0.438 &  1.59  & 
\\
\hline
\end{tabular}
\end{center}
Table II: Initial temperatures, transverse radii, and dissociation times  at 
different centralities for $J/\psi$ with $T_d \sim 1.1 T_c$ and 
number of flavour, $N_F = 0$. All the averages here are over the impact
parameter.

The survival probability, which is the ratio of bound states produced
by the escaped pairs relative to the number of bound states that would
be formed in the absence of QGP, is given by~\cite{chum}
\begin{equation}
S_Q(p_T,b) = \frac{3\,\int_0^{R_t}\,\phi_{\mathrm{max}}\,
\left[1 - \left(\frac{r}{R_t}\right)^2\right]^{1/2}r\,dr}{\pi R_t^2},
\label{eq18}
\end{equation}
where 
\bea
\phi_{\mathrm{max}}&=& \pi  :~~~~~~~~~~~~~~~~~~~ z \leq -1,\nonumber\\
                   &=& \cos^{-1}|z| : ~~~~~-1 \leq z \leq 1,\nonumber\\
                   &=& 0    :~~~~~~~~~~~~~~~~~~~ z \geq 1
\label{eq19}
\eea
with $z = [(R_s^2-r^2)m_R-\tau_{0f}^2p_T^2/m_R]/(2r\tau_{0f}p_T)$.
We assume a pure gluonic plasma with $T_c\,\sim\,270$ MeV~\cite{lat3}.

Suppression due to Debye screening continues till the temperature of the
QGP drops below $T_d$.  
The QGP starts hadronizing at $T_c$ (equivalently, at 
$\tau_Q=T_i^3\tau_i/T_c^3$,
the time when phase transition starts). During the
time interval $\tau_Q-\tau_d$, the remaining resonances will not be 
dissociated as long as $T_c$ is lower than $T_d$. During the mixed phase
interval ($\tau_H-\tau_Q$) QGP part does not contribute to the suppression for
$T_c < T_d$.
Here $\tau_H=g_Q\tau_Q/g_H$, where $g_Q$ and $g_H$ are the statistical
degeneracy for QGP and hadronic phases respectively.
However, the co-moving absorption starts at $\tau_Q$ and the density of
the co-moving hadrons is determined by the temperature $T_c$.
Following Ref.~\cite{gavin} we have checked that this effect is very small
in the present scenario.
Hence we do not consider suppression due to co-mover absorption here. 

Finally, the $p_T$-distribution of heavy resonances is given by 
\be
\frac{dN_{J/\psi}}{dp_T} = 
\frac{d\sigma_{J/\psi}^{NN}}{dp_T}
\frac{\int_0^{b_m}\,d^2b\,S_Q(p_T,b)
ABT_{AB}(b)}{\int_0^{b_m}\,d^2b\,\left[1-P_0(b)
\right]}
\label{eq23}
\ee

However, with nuclear absorption the above equation will be modified to: 
\be
\frac{dN_{J/\psi}}{dp_T} = 
\frac{\int_0^{b_m}\,\frac{d\sigma_{J/\psi}^{AB}}{d^2bdp_T}\,d^2b\,S_Q(p_T,b)\,
}{\int_0^{b_m}\,d^2b\,\left[1-P_0(b)
\right]},
\label{eq24}
\ee
Similar expression can be obtained for $\psi^{\prime}$ and $\chi_c$.

\section*{III. Results}
In Fig.~(\ref{fig1}) we compare our results  with CDF data where 
$m_c=1.3$ GeV, $Q^2=m_c^2+p_T^2$ have been used. It is 
clear  that the data is reproduced reasonably  well with the values of 
$F$ given in Table III. Any effect due to  $k_T$-broadening is ignored here.
However, it might be mentioned that the low $p_T$ data on $\Upsilon$ production
could be explained in color evaporation model by assuming substantial $k_T$
smearing \cite{schuler1}.
\begin{center}
\begin{tabular}{|lr|c|c|c|}\hline
&$F[J/\psi]$ & $F[\psi^{\prime}]$ &$\sum_J\,B(\chi_{cJ}\,\ra\,J/\psi\,X)\,F[\chi_{cJ}]$ & 
\\
\hline
& 0.045  & 0.0126 & 0.005 &
\\
\hline

\end{tabular}
\end{center}
\centerline{Table III: Values of the universal factor used in the calculation
(no feed down).
}
\vskip .1in

\bef
\centerline{\psfig{figure=fig1_rev.eps,height=8cm,width=10cm}}
\caption{
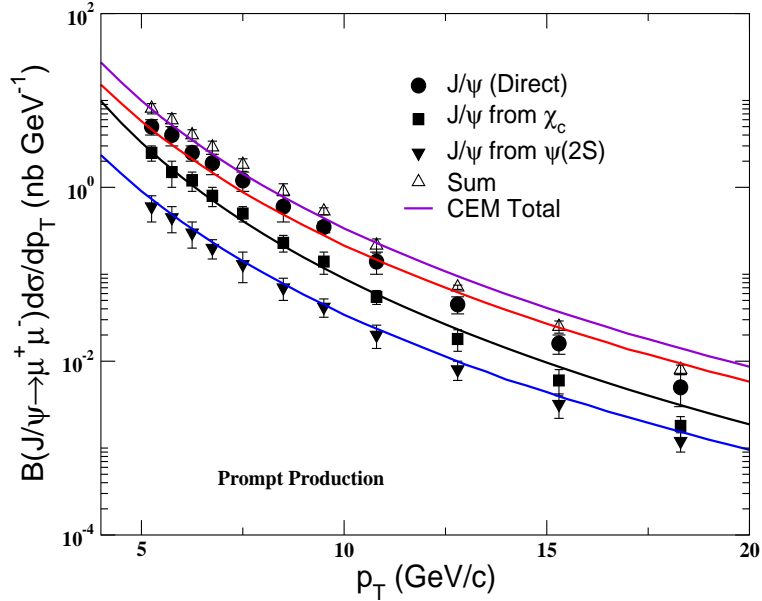
CDF data~\cite{cdf} for prompt $J/\psi$ 
 production is compared with CEM (LO plus gluon splitting) prediction 
at $\sqrt{s} = 1800$ GeV. CTEQ(LO) has been used in all the calculations. 
}
\label{fig1}
\eef

Next we consider the $J/\psi$ production at LHC energies. First of all, 
in Fig.~(\ref{fig2}), we show  $d\sigma/dp_T$ of $J/\psi$ and $\psi^\prime$
from $p-p$ collisions at $\sqrt{s} = 5500$ GeV. 
Here the yields
correspond to the direct productions {\it i.e.} feed down from higher
resonance states have been ignored with $m_c = 1.3$ GeV. 
The feed down contribution can be estimated by
multiplying the yield with appropriate branching ratio. 

\bef
\centerline{\psfig{figure=fig2_rev.eps,height=8cm,width=10cm}}
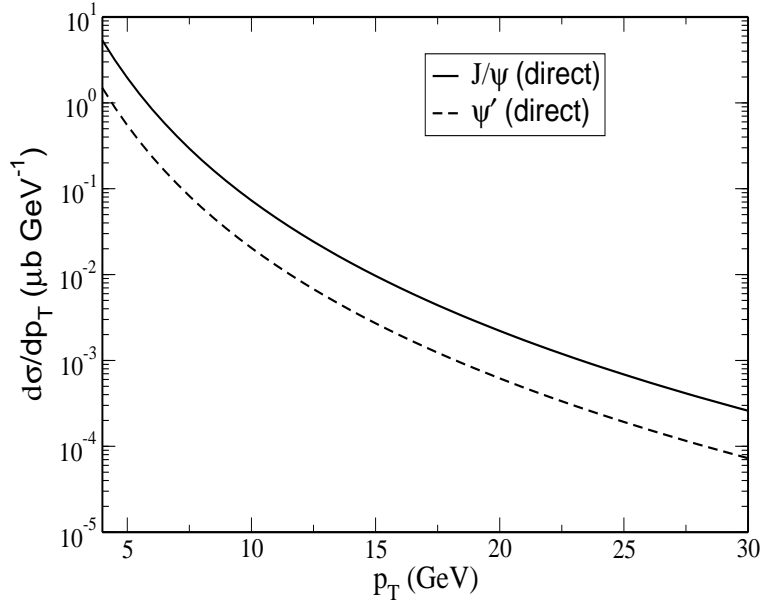
\caption{Differential cross-sections of $J/\psi$ and $\psi^{\prime}$
in $p-p$ collisions at $\sqrt{s} = 5500$ GeV. Only direct contributions
are shown.}
\label{fig2}
\eef

The  differential $p_T$ distributions of $J/\psi$ and $\psi^{\prime}$
from  $p-p$ scattering (as shown in Fig.~(\ref{fig2})) provide the 
baseline for $Pb-Pb$ collisions at LHC energies. To obtain
charmonium $p_T$ distributions at LHC energies we use eq.(\ref{eq23}) which
gives $p_T$ spectra without nuclear and
comover absorptions. 
The amount of nuclear absorption can be calculated using 
$\sigma_{\mathrm{abs}}=\sigma_{\mathrm{abs}}(\sqrt{s_0})\,
(s/s_0)^{\Delta/2}$, where $s_0=17.3$ GeV, $\sigma_{\mathrm{abs}}(s_0)=5\pm 0.5$
mb and $\Delta=0.125$~\cite{hardpro}. 
The nuclear absorption of heavy resonances at
SPS energies is well studied and at LHC energies it can be eliminated 
using $p-A$ collisions with the $\sqrt{s}$-dependent
cross-section mentoned above. In the present work, our motivation is to see
the anomalous suppression due to Debye screening, we do not include 
nuclear suppression in our calculation. We also note here that the
co-moving absorption will start when the plasma begins to hadronize. 
We have seen that the co-moving
absorption is minimal with the parameters used in Ref.~\cite{gavin}. The 
extrapolation of these
parameters to high energies is not straightforward.  
Therefore, in our calculation we do not include the hadronic
absorptions. The expected $p_T$ distributions of $J/\psi$ at LHC energies
are shown in Fig.~(\ref{fig3}) for $0 \leq b\leq 3$ fm
and $6 \leq b\leq 9$ fm
centralities with and without Debye screening.    
For $T_d \sim
1.1\,T_c$ we see that the $J/\psi$s with $p_T < $ 15(10) GeV are suppressed
for $0 \leq b \leq 3 (6 \leq b \leq 9)$ fm centrality. 
However, the other value of $T_d$ (see Table I)
the $J/\psi$s will not be suppressed for the same value of the hard fraction
$x$. We take  
$T_d = 1.5 T_c$ in Fig.~(\ref{fig4}). It is seen that 
suppression occurs below $p_T \sim 6$ GeV.
The value of the hard fraction $x$ is not known at LHC energies at this
stage. The
survival probability decreases as $x$ increases, since in that case,
the initial temperature will be high (see eqs.(\ref{eq15}) and (\ref{eq16}))
as demonstrated in fig.~(\ref{fig4a}).

   For $\psi^{\prime}$ the $T_d$ values are small compared to that of
$J/\psi$ and hence $\psi^{\prime}$ will be suppressed substantially. This
is shown in Fig.~(\ref{fig5}) with $T_d \sim 0.15 T_c$ at two
different centralities. It shows that even a very high $p_T$ ($\sim$ 30 GeV)
 $\psi^{\prime}$
is suppressed. We also plot the $p_T$ spectra of $\psi^{\prime}$ 
for average values of $T_d$ at three centralities in 
Fig.~(\ref{fig6}) to show the centrality dependence of heavy resonance 
suppression. We observe that the suppression is small in going from most
central to peripheral collisions as expected.  
\bef
\centerline{\psfig{figure=fig3_rev.eps,height=8cm,width=10cm}}
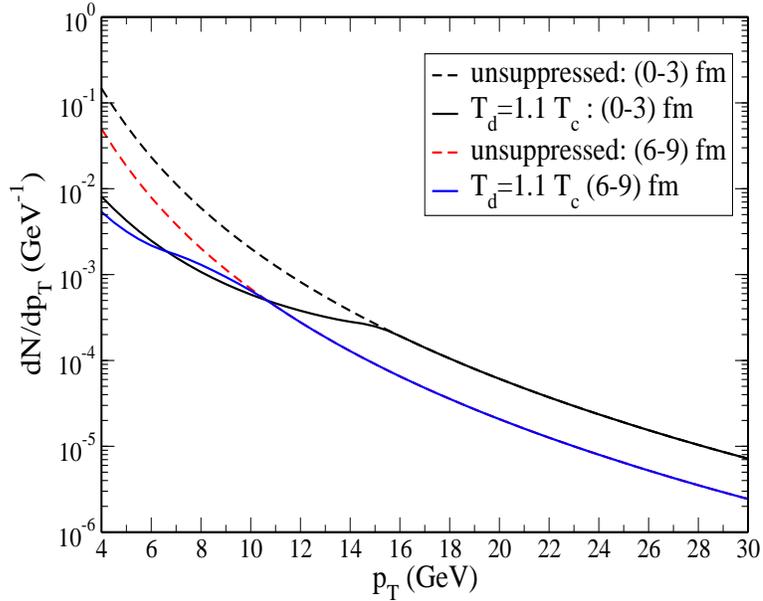
\caption{$p_T$ distribution of $J/\psi$ 
 in $Pb-Pb$ collisions at $\sqrt{s} = 5500$ GeV for $T_d = 1.1 T_c, x = 0.3$
and $\tau_i=0.5$ fm/c. 
}
\label{fig3}
\eef

\bef
\centerline{\psfig{figure=fig4_rev.eps,height=8cm,width=10cm}}
\caption{Same as Fig.~(\protect\ref{fig3}) for $T_d = 1.5 T_c$.} 
\label{fig4}
\eef

\bef
\centerline{\psfig{figure=fig3b_rev.eps,height=8cm,width=10cm}}
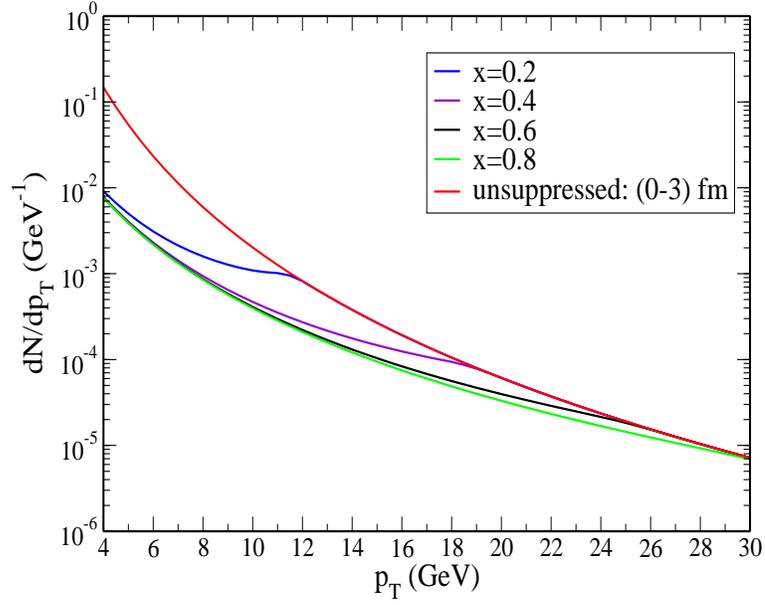
\caption{Same as Fig.~(\protect\ref{fig3}) for $T_d = 1.1 T_c$ for different
values of the hard fraction with impact parameter in the range
$0 \leq b \leq 3$ fm.} 
\label{fig4a}
\eef

\bef
\centerline{\psfig{figure=fig5_rev.eps,height=8cm,width=10cm}}
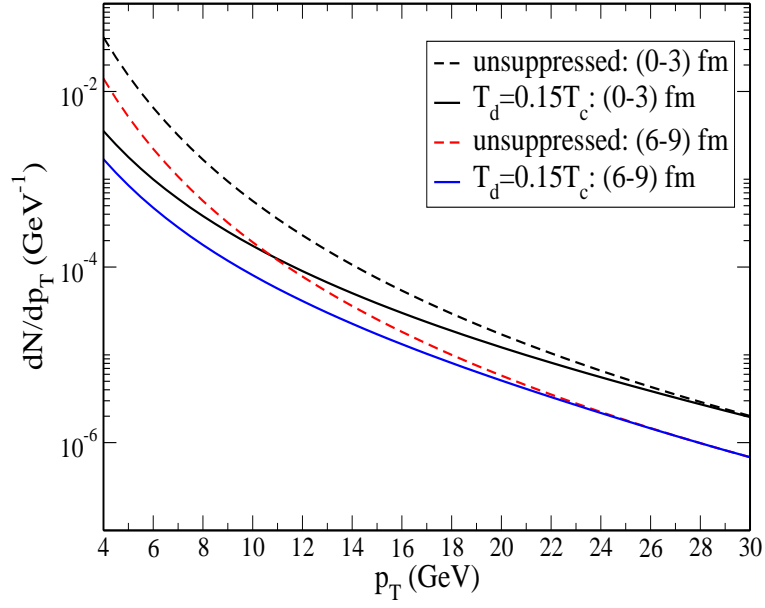
\caption{Same as Fig.~(\protect\ref{fig3}) for $\psi^{\prime}$ with 
$T_d = 0.15 T_c$.} 
\label{fig5}
\eef

\bef
\centerline{\psfig{figure=fig6_rev.eps,height=8cm,width=10cm}}
\caption{Same as Fig.~(\protect\ref{fig5})  with 
$T_d = 0.63 T_c$.} 
\label{fig6}
\eef

The prediction at RHIC energies is shown in Fig.~(\ref{fig7}). It is seen
that $J/\psi$ is not suppressed even for most central collisions as
the initial temperature is very close to the dissociation temperature.
However, $\psi^{\prime}$ is suppressed and the critical values of
$p_T$ depend on the the dissociation temperature used.
\bef
\centerline{\psfig{figure=fig7_rev.eps,height=8cm,width=10cm}}
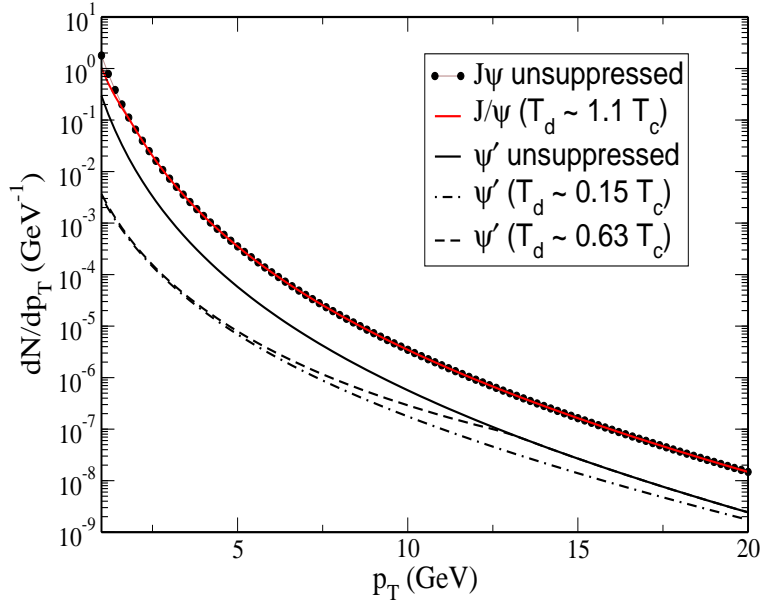
\caption{Same as Fig.~(\protect\ref{fig3}) at $\sqrt{s}= 200$ GeV, $\tau_i=
0.6 $ fm/c, $x=0.1$  and $0 \leq b \leq 3$ fm.} 
\label{fig7}
\eef

\section*{IV. Summary and discussions}
In this work we have calculated the $p_T$ distribution of charmonium
by using colour evaporation model ( LO with gluon splitting)
for RHIC and LHC energies. To validate our formalism we first
reproduce the CDF data for $p_T$ distribution of $J/\psi, \psi^{\prime},$
and $\chi_c$ using CEM. It is assumed that at LHC energies QGP is formed
and the heavy resonances are suppressed in a QGP due
to Debye screening. We have calculated the survival probability using
the prescription of Ref.~\cite{chum} and convoluted it with the $p_T$
distribution of various resonances. First, the $p_T$ spectra of various
charmonium states are obtained in nucleon-nucleon collision  
and then these are convoluted with
Glauber model to get the same in nucleus-nucleus
collisions. The centrality dependence of transverse momentum distribution
is also estimated. 
We have seen that $J/\psi$ with $p_T > 15 (10)  $ GeV will not be suppressed  
in  nuclear collisions for the impact parameter in the range
$0\leq b \leq 3$ fm ($6\leq b \leq 9$) fm.
We do not include the co-moving suppression due to hadrons,
as this is found to be small within the existing model parameters~\cite{gavin}.

\end{document}